\documentclass[
 reprint,
 amsmath,amssymb,
 aps,
 longbibliography,
  prl,
 citeautoscript
]{revtex4-1}

\usepackage{graphicx}
\usepackage{dcolumn}
\usepackage{bm}

\usepackage[english]{babel}
\usepackage{hyperref}
\usepackage{times}
\usepackage{amsfonts,amssymb,bm}
\usepackage{amsmath}
\usepackage{epsfig}
\usepackage{mathrsfs}
\usepackage{color,soul}
\usepackage{bbold}
\usepackage{lineno}
\usepackage{pstricks}
\usepackage{mathtools}
\usepackage{textcomp}
\usepackage{braket}
\usepackage{empheq}
\usepackage{epstopdf}
\usepackage{subfig}
\usepackage{amsmath}
\usepackage{amsthm}
\usepackage{dcolumn}
\usepackage{bm}
\usepackage[babel]{csquotes}
\usepackage[font=small]{caption}
\usepackage{mwe}

\newtheorem{theorem}{Theorem}
\usepackage[framemethod=default]{mdframed}
\usepackage[labelfont=bf,labelsep=quad,justification=raggedright]{caption}

\theoremstyle{definition}
\newtheorem{definition}{Definition}

\begin{document}

\title{Experimental two-way communication with one photon}

\author{Francesco Massa$^{1}$, Amir Moqanaki$^{1}$, \"Amin Baumeler$^{2, 5}$, Flavio Del Santo$^{1, 2}$, Joshua A. Kettlewell$^{3, 4}$, Borivoje Dakic$^{2}$, Philip Walther$^{1}$}

\affiliation{\normalsize{$^{1}$Vienna Center for Quantum Science and Technology (VCQ), Faculty of Physics, University of Vienna, Boltzmanngasse 5, Vienna A-1090, Austria}}
\affiliation{\normalsize{$^{2}$Institute for Quantum Optics \& Quantum Information (IQOQI), Austrian Academy of Sciences, Boltzmanngasse 3, Vienna A-1090, Austria}}
\affiliation{\normalsize{$^{3}$Singapore University of Technology and Design, 8 Somapah Road, Singapore 487372}}
\affiliation{\normalsize{$^{4}$Centre for Quantum Technologies, National University of Singapore, 3 Science Drive, Singapore 117543}}
\affiliation{\normalsize{$^{5}$Facoltà Indipendente di Gandria, Lunga Scala, 6978 Gandria, Switzerland}}

\begin{abstract}
Superposition of two or more states is one of the fundamental concepts of quantum mechanics and provides the basis for several advantages offered by quantum information processing. In this work, we experimentally demonstrate that quantum superposition allows for two-way communication between two distant parties that can exchange only one particle once, an impossible task in classical physics. This is achieved by preparing a single photon in a coherent superposition of the two parties' locations. Furthermore, we show that this concept allows the parties to perform secure and anonymous quantum communication employing one particle per transmitted bit. These important features can lead to the realization of new quantum communication schemes, which are simultaneously anonymous, secure and resource-efficient.
\end{abstract}

\maketitle

\section*{Introduction}
In recent decades, developments in the study of quantum information science have gained insights that promise to revolutionise the future of information processing. Among them, quantum communication is one of the earliest known applications demonstrating the clear advantage of quantum systems. The transmission of quantum states, in fact, allows for communication features that are not achievable with merely classical resources, such as information-theoretically secure quantum key distribution (QKD) \cite{BB84, E91, B92, QKD1, QKD2} or quantum secure direct communication (QSDC) \cite{rev2, QSDC6, QSDC7}.

In terms of efficiency, it has been demonstrated, both theoretically and experimentally, that quantum protocols reduce the information transfer required to perform some specific distributed computational tasks \cite{CC1, CC2, CC3, fin1, fin2, fin3, fin4}. Some of these schemes provide an exponential advantage with respect to their classical counterparts. At the same time, quantum systems allow for a decrease in the amount of physical resources necessary for communication \cite{DC1, DC2, DC3, RAC1, RAC2, RAC3}.

Along these lines, a recent theoretical result \cite{Bori} has shown that, by means of quantum superposition, it is possible to perform two-way communication between two distant parties that only exchange a single particle. Such an operation is impossible in classical physics, where two-way communication can be realised only if the parties exchange two particles, one per party, or if the same particle goes back and forth between them. Thus, for this specific task, quantum mechanics determines a reduction in number of particles to be used or, alternatively, in the time employed for the communication.

In this work, we experimentally demonstrate the two-way signalling via superposition of single photons.
Furthermore, we advance the scheme proposed in \cite{Bori} for performing QSDC. Our method achieves information-theoretically secure transfer of classical bits between two parties, given a shared single-particle superposition state.  With respect to other proposed QSDC schemes \cite{rev2}, our protocol has two advantages: 1) the direction of communication between the parties is hidden, as for quantum anonymous communication \cite{AQC1, AQC2, AQC3}, 2) security is not affected by multi-photon emission. We exploit the latter property to realize an implementation of the protocol that is robust against losses.

Our results show that a feasible quantum resource, such as superposition, allows for communication features that are classically impossible and can support the development of novel schemes.

\section*{Two-way signalling with a single photon}
\begin{figure}[t]
\centering
\includegraphics[width=\columnwidth]{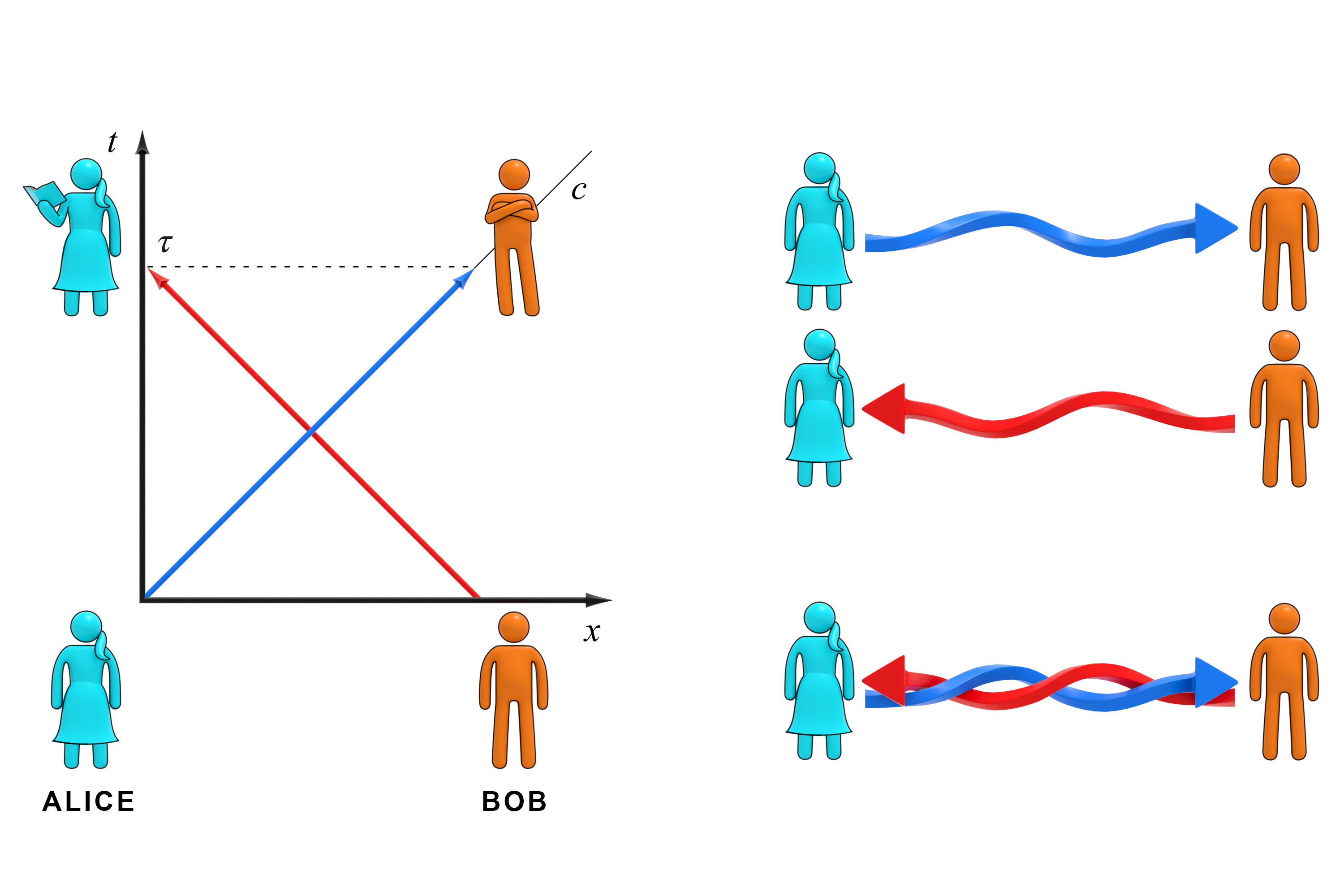}
\captionof{figure}{\footnotesize \textbf{Diagrams of communication between two distant parties.} Classically, a single carrier travelling with finite speed, bounded by the speed of light $c$, can transmit information either from Alice to Bob (blue arrow) or from Bob to Alice (red arrow) only, if the time $\tau$ allowed for the communication is shorter than the time the carrier takes to travel more than once the distance between Alice and Bob (space-time diagram on the left). An information carrier in quantum superposition permits to overcome this limitation and carry out a two-way communication process (scheme on the right).} 
\label{fig:alicebob}
\end{figure}

In order to show two-way signalling, we consider a communication game in which a referee respectively assigns two random input bits, $x$ and $y$, to two distant communication parties, named Alice and Bob, who are then allowed to exchange one particle. We call $\tau$ the time it takes for the exchange to be completed, that is the interval between the time at which the particle leaves Alice's or Bob's location and the time at which it is detected. We assume $\tau$ shorter than the time required to a physical object to travel more than once the distance between Alice and Bob (see figure \ref{fig:alicebob}). When the exchange is completed, the referee asks Alice and Bob to reveal two output bits, $a$ and $b$: they win the game if they both guess correctly the value of the other player's input (i.e. if $a=y$ and $b=x$). This game can be considered a variation of the well known ``guess your neighbour's input'' (GYNI) game \cite{GYNI}. Under the constraint that the parties can only exchange one particle within the time window $\tau$, only two possible causal relations between variables $x$, $y$, $a$ and $b$, are possible: either $x$ influences $a$ and $b$, whereas $y$ influences $b$ only (corresponding to a one-way communication from Alice to Bob) or $y$ influences $a$ and $b$, whereas $x$ influences $a$ only (one-way communication from Bob to Alice). Accordingly, the joint probability distribution $p(ab|xy)$ results in a classical mixture of the two one-way signalling distributions. This imposes a maximal probability value of 1/2 of winning the game \cite{branciard}.

Let us now consider the case of a single quantum particle prepared in a coherent superposition between Alice's and Bob's respective locations:
\begin{equation} \label{eq:prot1}
\vert \psi_{in}\rangle = \frac{1}{\sqrt{2}}({\widehat{a}}^{\dagger}+ {\widehat{b}}^{\dagger})\vert 0\rangle ,
\end{equation}
where ${\widehat{a}}^{\dagger}$ and ${\widehat{b}}^{\dagger}$ are the particle creation operators at Alice's and Bob's location, respectively, and $\vert 0 \rangle$ is the vacuum state.\\
Alice and Bob encode the bits $x$ and $y$ in the phase of the particle, obtaining the state:
\begin{equation} \label{eq:prot2}
\vert \psi_{encode}\rangle = \frac{1}{\sqrt{2}}((-1)^x \ {\widehat{a}}^{\dagger} + (-1)^y \ {\widehat{b}}^{\dagger})\vert 0\rangle.
\end{equation}
A 50/50 beam splitter is placed at the centre of the path between Alice and Bob. The action of the beam splitter can be expressed by the following transformations:
\begin{eqnarray} \label{eq:prot3}
{\widehat{a}}^{\dagger} \longrightarrow \frac{1}{\sqrt{2}}({\widehat{a}}^{\dagger} + {\widehat{b}}^{\dagger}), \\ 
{\widehat{b}}^{\dagger} \longrightarrow \frac{1}{\sqrt{2}}({\widehat{a}}^{\dagger} - {\widehat{b}}^{\dagger}).
\end{eqnarray}
Due to interference, after the device the final state of the photon is:
\begin{equation} \label{eq:prot4}
\vert\psi_{fin}\rangle=
\begin{cases}
\ \ \ {\widehat{a}}^{\dagger} \vert 0 \rangle, & \text{if $x = 0$ and $y = 0$,} \\
\ \ \ {\widehat{b}}^{\dagger} \vert 0 \rangle, & \text{if $x = 0$ and  $y = 1$,} \\
-{\widehat{b}}^{\dagger} \vert 0 \rangle, & \text{if $x = 1$ and $y = 0$,} \\
-{\widehat{a}}^{\dagger} \vert 0 \rangle, & \text{if $x = 1$ and $y = 1$.}
\end{cases}
\end{equation}
This means that, by checking whether they detect the particle or not, Alice and Bob can infer the parity, $r$, of $x$ and $y$. This piece of information, combined to the knowledge of their input bits, allows them to ideally win the game with probability $1$, thus showing genuine two-way communication. 
\vspace{14pt}

\section*{Application for anonymous communication}
We now present a secure two-party quantum communication protocol that provides anonymity of the direction of communication between Alice and Bob. This is achieved via implementation of the previously described two-way communication scheme as a primitive. As communication via the protocol is two-way, we convert this to a direct message system by allowing only one party to transmit a message at a time, and the other to transmit only random bits. 

Our protocol makes an implicit assumption that Alice and Bob share a quantum channel and many copies of the required superposition state, $\ket{\psi_{in}}$, which is known to be a powerful recourse for secure communication \cite{entanglementresource}. Such states could be supplied on demand via a trusted server assuming the channel between the server and Alice, and the server and Bob, are secure to a possible eavesdropper. Alternatively, they could be in theory produced and stored by the two parties when they meet and then used at a later moment. Prior to the protocol, each state $\ket{\psi_{in}}$ shared between the parties is labelled with index $i$.  

For each round of communication, $i$, both Alice and Bob perform local phase operations $\ket{\psi_{encode}}_i=  \frac{1}{\sqrt{2}}((-1)^{x_i} \ {\widehat{a}}^{\dagger} + (-1)^{y_i} \ {\widehat{b}}^{\dagger})\ket{0}$ to encode bits $x_i$ and $y_i$, respectively. Both send their part of the state $\ket{\psi_{encode}}_i$ via the quantum channel and detect any returning photon. Detection of a photon reveals the parity bit $r_i=x_i \oplus y_i$ to each party. Assuming Alice wishes to send an $M$-bit message $\{X_1, \dots , X_M\}$ to Bob, the protocol can be described by the following sequence of steps:

\begin{enumerate}
\item \textbf{Decline communication.} If no message is to be sent, Alice and Bob select the bits $x_i$ and $y_i$ uniformly at random. 

\item \textbf{Declaration of the communication direction.}  Alice initializes communication via setting  $x_i=1$ for $d$ iterations of the protocol, where $d$ is chosen as to be sufficiently large as to be sufficiently improbable to occur by chance. Detection of $d$ repeated $x_i = 1$ results by Bob indicates that Alice intends to send a message. Should Bob simultaneously declare his intention to communicate, the protocol is aborted.

\item \textbf{Transmission of the message.} Alice sets $x_i$ = $X_i$, for $i$ going from $1$ to $M$. Bob may or may not detect a photon, thus obtaining the parity value $r_i = y_i \oplus X_i$, from which the bit $X_i$ can be deduced. 

\item \textbf{Declaration of the end of the message.} To end the message transmission, Alice sends $x_i =0$ for $d$ iterations of the protocol. Alice and Bob return to step 1.

\end{enumerate}

The description of the scheme makes no assumption on the power of an eavesdropper, allowing for information-theoretic security. Interception of a photon between the two parties will leak exactly the parity between $x_i$ and $y_i$, given by the position of the photon after the interference at the central beam splitter. This may be observed as the four possible states of $\ket{\psi_{encode}}$ form two pairs that are identical under global phases, which cannot be observed via measurement. As such, only a  single bit of information may be obtained by an eavesdropper. As each bit $y_i$ is chosen uniformly at random, this bit contains no information about $x_i$, provided that $y_i$ is unknown, and thus leaks no information regarding Alice's message bit $X_i$. Bob's input thus acts as a random one-time  pad. As communication is two-way, pad bits $y_i$ are also obtained by Alice, and as such the scheme is anonymous in the direction of the message and pad. A detailed security analysis of the protocol is provided in appendix E.

Such a system may be easily altered to become resistant to experimental losses within a realistic implementation. Losses caused by an erasure channel may result in no photon being detected by Bob when required, causing a single bit error in the received message. Additional errors may be caused by imperfections in the experimental set-up, such as dephasing or non-optimal interference visibility. However, errors can be overcome, without compromising security, by adding redundancy to the protocol, as discussed in appendices D and E. Experimentally, this type of error correction requires no fast switching elements if channel losses are high, greatly simplifying practical applications.

In contrast to other quantum communication schemes, the security of the protocol is preserved even in the case of simultaneous multi-photon emission from the source, as shown in appendix E (Theorem 1). This underlines the feasibility of our protocol when using realistic single-photon sources.

Although anonymous communication may be performed between two parties via use of shared classical data, our method demonstrates the power of a superposition state as a resource for communication. 

\section*{Experimental Results}
\subsection*{Implementation of the communication game}
\begin{figure}
\centering
\includegraphics[width=\columnwidth]{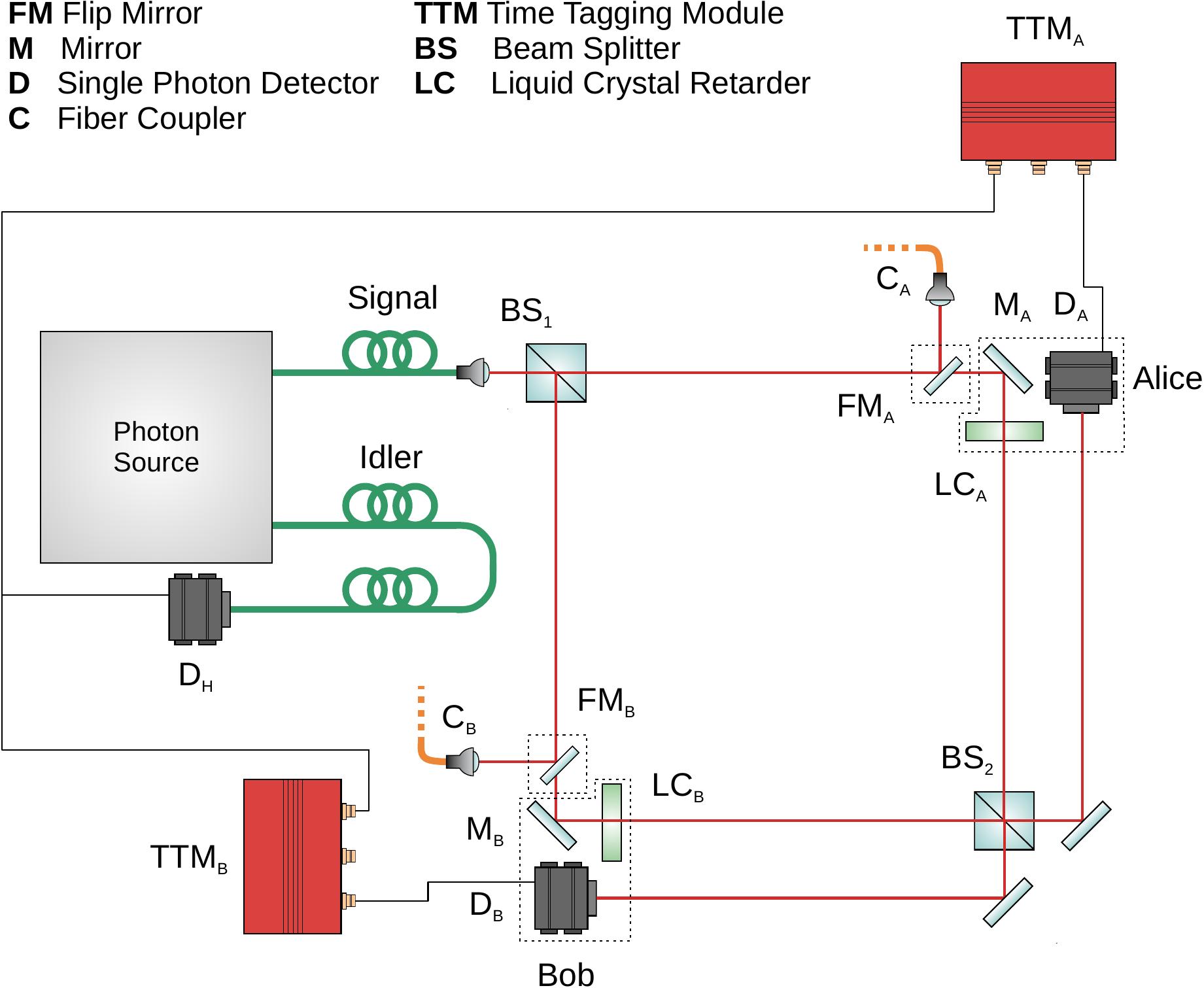}
\captionof{figure}{\footnotesize \textbf{Experimental set-up.} Single-photon pairs are produced through spontaneous parametric down-conversion (SPDC). For each pair, one photon is used to herald the presence of the other one, which is sent to a Mach-Zehnder interferometer. Alice and Bob occupy the area around the mirrors M$_{\text{A}}$ and M$_{\text{B}}$, where, for each of them, a liquid-crystal phase shifter, for phase encoding, and a photon detector are placed. After the second beam splitter, the photons can travel to Alice or Bob, according to the parity of the input bits. Removable mirrors are used to measure the time at which Alice and Bob receive the photons from the source for the purposes explained in the main text. These mirrors steer light to fibers that can be connected to either Alice's or Bob's detector. For more details about the set-up, we refer to appendix A.
}
\label{fig:setup}
\end{figure}

The set-up for the implementation of the communication game is shown in figure \ref{fig:setup}. A heralded single photon is sent to one of the input ports of a first beam splitter, which puts the photon in a superposition state between Alice's and Bob's locations. Then, Alice and Bob encode their bits in the phase of the photon and direct it to a second beam splitter, which creates the final state $\vert \psi_{fin}\rangle$. This scheme represents a Mach-Zehnder interferometer. 

\begin{figure}
\centering
\includegraphics[width=0.95\columnwidth]{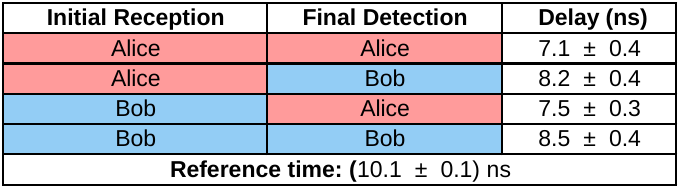}
\captionof{table}{\footnotesize \textbf{Time measurement results.} The four possible delays between the initial reception and the final detection of the photon at Alice or Bob are shown in the table. They are compared to the time the photon would take to travel twice the minimum distance between the two parties, roughly equal to the diagonal of the interferometer, at the speed of light in vacuum (reference value). For each delay, the measurements are taken by unblocking only the corresponding path and recording the arrival-time statistical distributions for reception and final detection, respectively. The uncertainty on the delays are obtained from the standard deviations of the associated arrival-time distributions, dominated by the time jitter of our detectors. The uncertainty on the reference value is not statistical and is determined by the measurement of the minimum distance between Alice and Bob.}
\label{tab:times}
\end{figure}

In order to prove that each photon cannot be exchanged more than once between the two parties, we measure the delay between two events: the reception of the photon before the encoding and the final detection after the second beam splitter. Actually, there are four delays to be measured, according to whether the initial reception and the final detection of the photon are considered at Alice or Bob. The delays are slightly different due to the fact that the implemented interferometer is rectangular. The results of these measurements are shown in table \ref{tab:times}. It can be seen that, in all the cases, the time $\tau$ necessary for the photon exchange to be completed is shorter than the time the photon would take to travel twice the minimum distance between Alice and Bob (reference time) by more than three standard deviations. This excludes the possibility that the photon travels back and forth between Alice and Bob with less than $1 \%$ risk. More details about the adopted measurement method and the data analysis can be found in appendix B.

\begin{figure}
\centering
\includegraphics[width=\columnwidth]{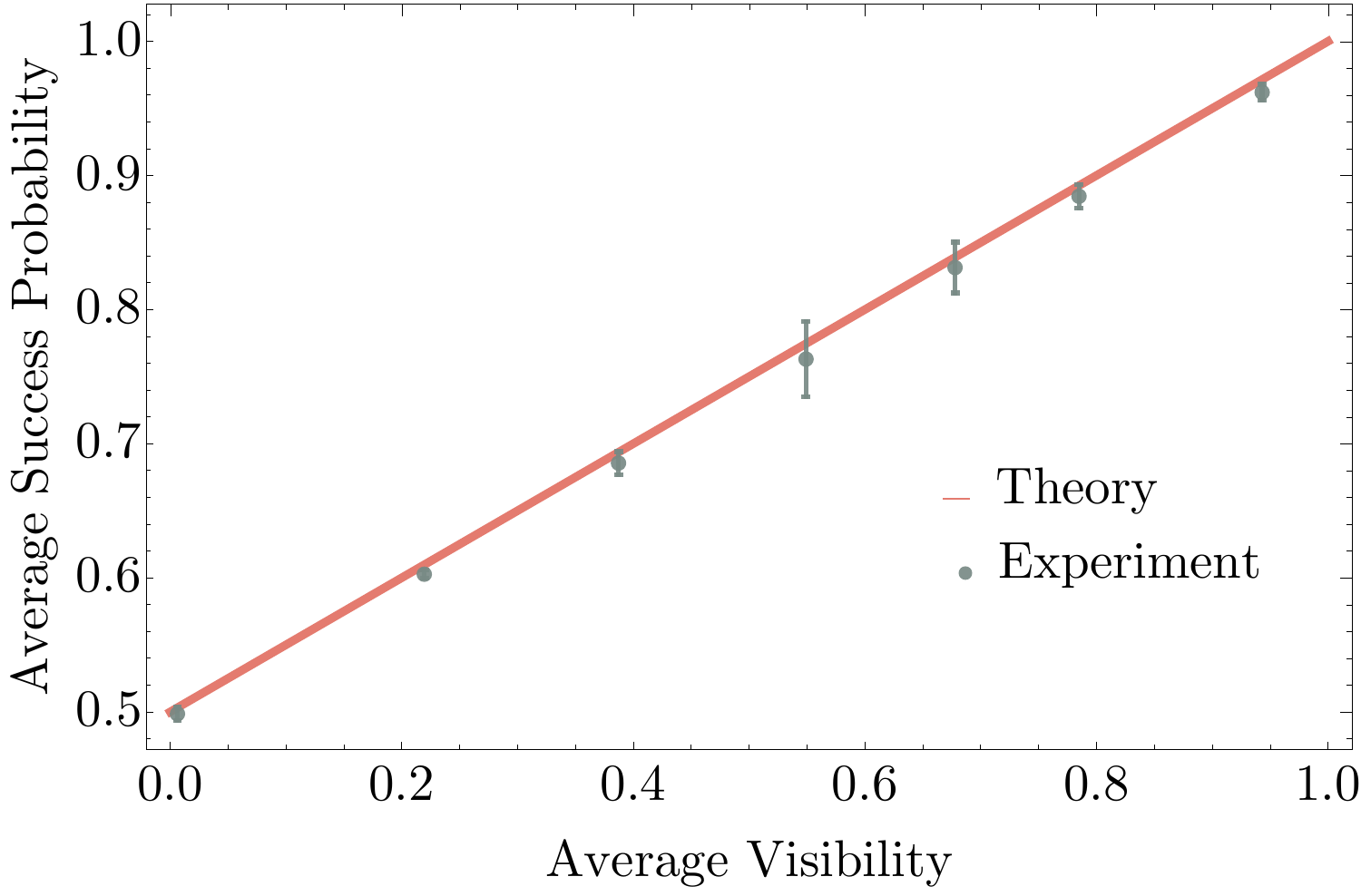}
\captionof{figure}{\footnotesize \textbf{Success probability vs interferometric visibility.} The plot shows the behaviour of the probability of winning the game with respect to the quality of the single-photon interference produced by the state Alice and Bob share, which is quantified by the average interferometric visibility. The visibility is varied by delaying one interferometric path with respect to the other: at zero visibility the two photon wave packets travelling in the two arms no longer overlap at the final beam splitter and the interference is completely cancelled. The equation of the red theoretical curve is $y = 0.5 (x + 1)$. The error on each probability is the standard error on the mean, obtained from the statistical variation over the sequence of input bits. For each point in the plot, a different random input sequence of bit pairs is generated.}
\label{fig:data}
\end{figure}

We estimate the probability of winning the game by using a random sequence of $100$ input bit pairs, one every $0.5$ s. In this time interval, we register an average number of photon detections of about $15\times 10^3$. For each bit pair, therefore, we compute the probability of success by counting how many photons go to the ``right'' output. We  then average the probability over the input sequence. Figure \ref{fig:data} shows the measured success probability for different values of the interferometric visibility in our Mach-Zehnder, averaged over the two output ports. The visibility at each port is defined as $(N_{MAX} - N_{MIN})/(N_{MAX}+ N_{MIN})$, where $N_{MAX}$ and $N_{MIN}$ are the maximum and minimum number of detections at that port. The success probability surpasses the classical limit as soon as the visibility is greater than zero.
For our maximally achieved visibility of $0.941 \pm 0.007$, we observe the maximal success probability of $0.961 \pm 0.006$.  At zero visibility the success probability is $0.498 \pm 0.006$, comparable with the maximal achievable value in the classical case ($0.5$). At this point, the effect of the quantum superposition is totally nullified.

In order to claim implementation of a two-way communication protocol with a single particle, we are required to demonstrate that Alice and Bob cannot share two or more photons at the same time. This can be shown by measuring the heralded second-order correlation function at zero delay of our photon-pair source, $g^{(2)}(0)$ \cite{g2}. This is a number between $0$ and $1$, quantifying the amount of multi-photon emission from the source. A value of $g^{(2)}(0)$ closer to $1$ would imply that two or more photons are sent simultaneously to the interferometer. For an ideal heralded single-photon source this number is 0. We measure  $g^{(2)}(0) = 0.004 \pm 0.010$, which is statistically compatible to $0$ and in line with the lowest values obtained in quantum optics experiment \cite{g2comp}. For more details about how this value was obtained we refer to appendix C.

\subsection*{Implementation of the anonymous communication protocol}
We experimentally demonstrate an application of the anonymous communication protocol with a probabilistic photon source and majority voting error-correction procedure. Although  the distribution of the superposition states is performed here by a trusted third party, our method allows us to anonymously send an image between two parties in a manner secure to any eavesdropper acting only between them.  
 
\begin{figure}
\centering
\includegraphics[width=\columnwidth]{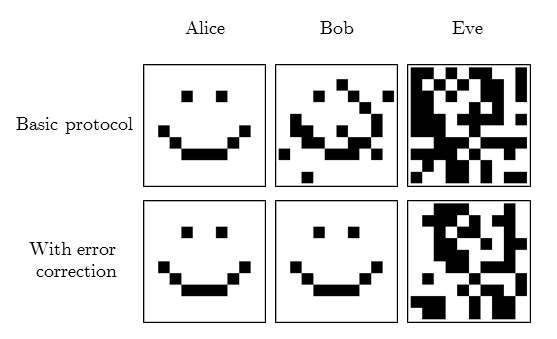}
\captionof{figure}{\footnotesize \textbf{Example of secure communication.} An example in which Alice sends a message in the form of a figure and Bob a random sequence with the same length is presented. The three columns report, in the order, the figure sent by Alice, the figure received by Bob and the parity of the bits sent by Alice and Bob, the only piece of information an eavesdropper, Eve, can obtain from the superposition state. Two cases are shown: the basic protocol, where each bit pair is sent once with an average probability of success of $88 \%$ and the error-corrected protocol, where each bit pair is sent five times, with an average probability of success of $100 \%$.}
\label{fig:images}
\end{figure}

The communication protocol employs the set-up described in the previous section and depicted in figure \ref{fig:setup}. A communication interval of $0.5$ s is set for each pair of bits $x_i$ and $y_i$, and the source emission rate is reduced so as to have an average of approximately three detection events per communication interval. Here we consider the sum of the detections Alice and Bob record. If Alice (Bob) receives one or more photons during a given communication interval, she (he) infers that  $r_i = 0$ ($=1$), and $r_i = 1$ ($=0$) otherwise. In appendix E we demonstrate security of the protocol under such conditions. The reader should note that, as the experimental setup heralds the photon in the interferometer with a non-number-resolving detector, potential higher-order-emission terms from the source are statistically mixed. As discussed in appendix E, this does not affect security.
Errors occur if in a given interval at least one photon goes to the ``wrong'' output or when no photon is detected by both Alice and Bob at the end of the interval. Such errors can be minimized by suitably choosing the average number of detections per interval (see appendix D for more details). When the error probability per bit, $p_b$, is lower than $50 \%$, the majority-voting error-correction code can be applied to further increase the success probability of the protocol. It consists in repeating the same message bit over $N$ communication intervals and selecting the outcome that occurs more often. We provide an example by implementing simple schemes with $N = 3$ and $N = 5$. The average success probability of the communication protocol without majority-vote procedure, measured by counting the successful transmission events for different random sets of $100$ bit pairs, is $0.88 \pm 0.01$. By implementing the error-correction schemes with three and five repetitions per bit pair, we obtain success probabilities of $0.93 \pm 0.01$ and $1.00 /- 0.01$, respectively.\\
We report an example where Alice sends a $10$ pixels $\times$ $10$ pixels image in black and white, corresponding to $100$ bits, and Bob sends a sequence of $100$ random bits. Figure \ref{fig:images} shows the outcome of the communication both for the basic protocol and for the error-corrected one with five repetitions per bit pair. 

\section*{Discussion}
We have experimentally demonstrated that, by using quantum superposition, it is possible to perform two-way communication between two parties that exchange only a single photon. The possibility that the photon travels back and forth between them or that two or more photons are simultaneously used is strictly ruled out by our implementation. Furthermore, we have designed and implemented a protocol for anonymous messaging via utilising two-way communication as a primitive, and shown that the method achieves information-theoretic security while being not compromised by photon losses and multi-photon emission. The security of our protocol is based on the impossibility of measuring the global phase for single photons. We, therefore, show a novel utilization of basic quantum-mechanical phenomena for communication.
Future developments hold the promise to approach real-world applications based on the recent progress in  bright deterministic single-photon sources \cite{quantumdots}, fast and low-loss optical switches \cite{switch} and high-efficiency single-photon detection \cite{nanowire}.
Recently, QKD schemes based on phase encoding  have raised great interest in the quantum cryptography community, because of providing a secure key rate that scales with the square root of the communication channel transmission \cite{TFQKD, PMQKD}. These protocols present some connections with ours, as they also employ relative-phase detection through first-order interference. However, their purpose is different from direct communication and they make use of coherent states, meaning that security is not based on the properties of fixed-photon-number states. Nevertheless, both lines of research show the importance of finding alternative communication schemes that can be advantageous over the current ones. From this perspective, examining in depth the relation between the two works can be beneficial for future investigations. 
\vspace{14pt}

\section*{Acknowledgements} 
We would like to thank Marcus Huber, Tommaso Demarie and Tiago Batalh\~{a}o for useful discussions. P.W. acknowledges support from the European Commission through QUCHIP (No. 641039) and ErBeStA (No. 00942), and from the Austrian Science Fund (FWF) through CoQuS (W1210-4) and NaMuG (P30067-N36), the U.S. Air Force Office of Scientific Research (FA2386-17-1-4011), and Red Bull GmbH. \"A.B. acknowledges the Swiss National Science Foundation (SNSF) under grant 175860, and the
Erwin Schr\"odinger Center for Quantum Science \& Technology (ESQ). F.D.S. acknowledges acknowledges financial support through a DOC Fellowship of the Austrian Academy of Sciences. J.A.K. acknowledges support from the Singapore National Research Foundation (NRF-NRFF2013-01). 

\vspace{0.5cm}

\section*{Appendix A: The experimental set-up}
\subsection*{The single-photon source}
We use an SPDC-based single-photon source in a Sagnac configuration \cite{sagnac}, with a 20-mm-long periodically-poled potassium tytanyl phosphate (PPKTP) crystal.\ The Sagnac loop is realized using a dual-wavelength polarizing beam splitter and two mirrors. The crystal converts a photon at $395$ nm into two photons at $790$ nm and orthogonal polarizations. The produced photons are coupled into single-mode fibers: one of them is sent to the Mach-Zehnder interferometer and the other is directly sent to a silicon avalanche photo-diode (APD) for heralding the presence of its twin in the interferometer. The use of polarizers for both photons of each pair ensures that a defined polarization state is produced, in particular $\vert H \rangle \vert V\rangle$, where H stands for ``horizontal'' and V for ``vertical'').

\subsection*{The interferometric set-up}
The interferometer we built is depicted in figure 1 in the main text. The distance between the mirror M$_{\text{A}}$ and the beam splitter BS$_2$ is $(106 \pm 1)$ cm, whereas the distance between M$_{\text{B}}$ and BS$_2$ is $(119 \pm 1)$ cm.\ The minimum distance between the regions occupied by Alice and Bob is the distance between the sides of the liquid-crystal phase shifters, equal to $(156 \pm 1)$ cm. The geometry of the interferometer is chosen so as to maximize the difference between the time photons take to travel from mirrors M$_{\text{A}}$ and M$_{\text{B}}$ to the detectors and the time they would take to travel twice the minimum distance between Alice and Bob, given the limits of space on our optical table. The two flip mirrors FM$_{\text{A}}$ and FM$_{\text{B}}$ are placed at $10.0$ cm of distance from M$_{\text{A}}$ and M$_{\text{B}}$, respectively. They are used to steer light to two fiber couplers, C$_{\text{A}}$ and C$_{\text{B}}$, connected to two $2$-m-long multi-mode fibers. The coupling in the multi-mode fibers is about $96 \%$. The distance between the flip mirrors and the couplers is also $10.0$ cm, so that photons reach the fibers at the same time they would arrive at M$_{\text{A}}$ and M$_{\text{B}}$ if FM$_{\text{A}}$ and FM$_{\text{B}}$ were not inserted. The uncertainty on all these distances is estimated to be $0.5$ cm. The detectors D$_{\text{A}}$ and D$_{\text{B}}$ are silicon APDs. We use them in a free-space configuration for the final detection of the photons but we connect them to the fibers from the couplers for the acquisition of the photon arrival-time distributions at the mirrors M$_{\text{A}}$ and  M$_{\text{B}}$. All the arrival times are measured by means of two different time-tag logic units, one for each detector, and are always referred to the detection of the heralding photon, used as a trigger.

The interferometer is passively stabilized by thermal and vibrational isolation so that the phase between the two arms is stable for about one minute. After this time, the phase can be re-set by means of a piezo actuator mounted in a trombone delay line, which can be used to delay one arm with respect to the other and therefore to change the interference visibility.
We re-set the piezo every $50$ input bit pairs, corresponding to about $25$ s. There are still some residual fluctuations of the phase around the stability point in this time interval, which, together with the standard poissonian fluctuations in the number of counts, determine the errors on the success probabilities reported in the main text.

The polarization of the photons entering the interferometer is set to ``horizontal'' (H), that means parallel to the optical table, before BS$_1$ by means of two waveplates and a polarizer. The slow axes of the two liquid-crystal phase shifters are aligned to the photon polarization. The refractive index along these axes depends on the voltage applied to the liquid crystal. We characterize the phase-shift with respect to the voltage and set a phase-shift of $0$ to encode the bit ´´0'' and of $\pi$ to encode the bit ´´1''.

\section*{Appendix B: Analysis of the photon arrival-time distributions}
Let us call $\Delta t_{\text{AB}}$ the time photons take to travel from the mirror M$_{\text{A}}$ to the detector D$_{\text{B}}$ along the arms of the interferometer. In the same way we call $\Delta t_{\text{AA}}$, $\Delta t_{\text{BA}}$, $\Delta t_{\text{BB}}$ the time photons take to go from M$_{\text{A}}$ to D$_{\text{A}}$, from M$_{\text{B}}$ to D$_{\text{A}}$ and from M$_{\text{B}}$ to D$_{\text{B}}$, respectively. The procedure to measure $\Delta t_{\text{AB}}$ is the following:
\begin{enumerate}

\item we block all the possible paths for the photons except for that one going from M$_{\text{A}}$ to D$_{\text{B}}$.

\item For each photon pair, we register the delay between the detection of the heralding photon and the detection of its correlated photon at D$_{\text{B}}$, after it travels through the interferometer. In this way we acquire the arrival-time distribution for the final detection at D$_{\text{B}}$, referred to the herald detection.

\item We turn up the flip mirror FM$_{\text{A}}$ and connect the multi-mode fiber from the coupler C$_{\text{A}}$ to the detector D$_{\text{B}}$. After correcting for the delay introduced by the fiber, we acquire the arrival-time distribution at M$_{\text{A}}$, as in point 2. 

\item We fit the two obtained distributions with gaussian functions and, for each of them, we consider the mean value and the standard deviation.

\item We calculate $\Delta t_{\text{AB}}$ as the difference of the mean values of the two distributions. Since the detections take place at the same detector and we use the same time-tag unit, the difference is not affected by further electronic delays. The error on $\Delta t_{\text{AB}}$ is obtained by adding in quadrature the standard deviations of the two distributions.
\end{enumerate}

For the measurement of $\Delta t_{\text{AA}}$, $\Delta t_{\text{BA}}$ and $\Delta t_{\text{BB}}$, we follow analogous procedures. In order to correct the delays introduced by the multi-mode fibers, their length is measured with a fiber-meter. We obtain $(2.080 \pm 0.004)$ m and $(2.088 \pm 0.004)$ m for the fibers connected  to C$_{\text{A}}$ and  C$_{\text{B}}$, respectively. The refractive index of the core, made of pure silica, is taken from \cite{index}. The errors on the fiber lengths and on the refractive index are negligible with respect to the standard deviations of the arrival time distributions. Supplementary Figure 1 shows the acquired arrival-time distributions, together with the related gaussian fits. 

Our detectors are single-photon counting modules from Excelitas, model SPCM-AQRH. This model has a typical jitter time (standard deviation) of $0.149$ ns. Since each peak is obtained by coincidence detection between two modules, if we consider only the effect of the jitter, we expect a standard deviation of $0.210$ ns. This value is compatible with those obtained for fiber-coupled detection but significantly lower than those obtained in case of free-space detection. We ascribe the mismatch to the imperfect alignment of the beam in the case of free-space detection.

As it can be seen from table 1 in the main text, the quantity $|\Delta t_{\text{X}} - \Delta t_{\text{ref}}|/\sigma_{\text{X}}$ is always above $3$, where $\Delta t_{\text{comp}}$ is the reference value and X can be AA, AB, BA or BB, respectively. This allows us to claim with less than $1 \%$ of risk that the two values are not compatible and then that the time the photons take to go from the mirror M$_{\text{A}}$  or M$_{\text{B}}$  to the detector is shorter than the time they would take to travel twice the minimum distance between Alice and Bob. In this procedure of comparison we neglect the error on $\Delta t_{\text{comp}}$ because the corresponding relative error is more than $4$ times lower than that on any $\Delta t_{\text{X}}$.

\section*{Appendix C: Measurement of the second-order correlation function at zero delay}
In order to measure the heralded second-order correlation function at zero delay, $g^{(2)} (0)$, we steer the photons to the couplers C$_{\text{A}}$ and  C$_{\text{B}}$, by means of the flip mirrors FM$_{\text{A}}$ and  FM$_{\text{B}}$, and we connect the related fibers to the detectors D$_{\text{A}}$ and  D$_{\text{B}}$. We consider the two-fold coincidence rates between the detection of the heralding photon and the detection of its correlated photon at D$_{\text{A}}$ or  D$_{\text{B}}$, which we call respectively CC$_{\text{HA}}$ and  CC$_{\text{HB}}$, and the three-fold coincidence rate, CC$_{\text{HAB}}$. We set the delays between the detections electronically in order to maximize CC$_{\text{HA}}$ and CC$_{\text{HB}}$ and in these conditions we evaluate $g^{(2)} (0)$, according to the following formula(\cite{formula}):

\begin{align}\label{eq:g2}
g^{(2)} (0) = \frac{2 \times \text{C}_{\text{H}} \times \text{CC}_{\text{HAB}}}{(\text{CC}_{\text{HA}} + \text{CC}_{\text{HB}})^2},
\end{align}
where $\text{C}_{\text{H}}$ is the single-count rate for the heralding photons.

We average the rates over $3$ minutes and obtain $g^{(2)} (0) = 0.004 \pm 0.01$, where the error is calculated from poissonian uncertainty on the count rates. This value is measured for $7$ mW of pump power in the source.

\section*{Appendix D: Error correction}
The error-correction procedure presented in the main text consists in repeating the encoding of each bit $N$ times, with $N$ odd, and performing majority voting, meaning that the result occurring at least $\frac{N + 1}{2}$ times is chosen. If the probability of error per bit is $p_b$, the overall probability of error after $N$ repetitions, $p_e$, is given by:
\begin{align}\label{eq:totalerror}
p_e = \sum\limits_{k = \frac{N+1}{2}}^{N} {N\choose k} p_b^k (1 - p_b)^{N - k}.
\end{align}

\begin{figure} 
\includegraphics[width=\columnwidth]{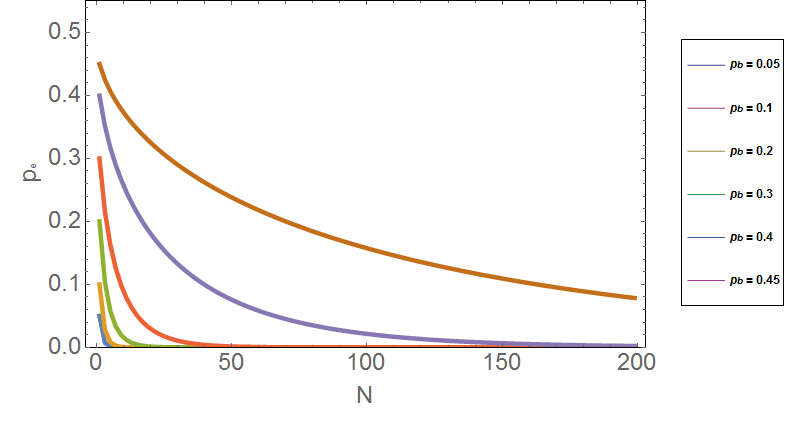}
\captionof{figure}{\footnotesize \textbf{$p_e$ vs $N$}. The plot shows the overall error probability after $N$ repetitions for different values of $p_b$. The slope of the curve reduces as $p_b$ approaches $0.5$.
}
\label{sfig:trend1}
\end{figure}

In figure \ref{sfig:trend1} we show the behaviour of $p_e$ with respect to $N$ for different values of $p_b$. It is clear that the higher is $p_b$, the slower $p_e$ goes to $0$. When $p_b = 0.5$, $p_e$ is independent of $N$ and for $p_b > 0.5$ the majority-voting procedure only worsens the probability of success.

\begin{figure}
\centering
\includegraphics[width=0.8\columnwidth]{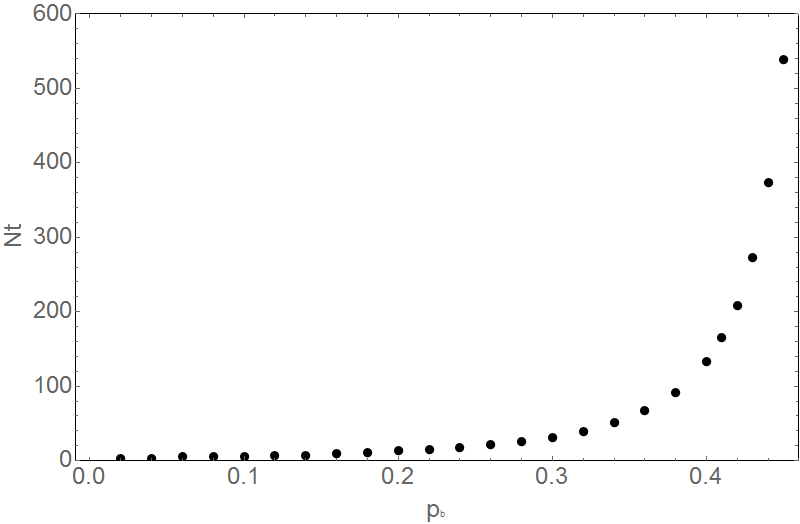}
\captionof{figure}{\footnotesize \textbf{$N_t$ vs $p_b$}. For this plot the threshold is set to 0.01. It can be noticed that $N_t$ starts increasing significantly for $p_b > 0.3$.
}
\label{sfig:trend2}
\end{figure}

We also consider the minimum amount of iterations $N_t$ necessary for $p_e$ to be below a threshold value $t$. In figure \ref{sfig:trend2}, we plot the dependence of $N_t$ on $p_b$ for $t = 0.01$. This trend is shown in figure 3. We can observe that $N_t \rightarrow \infty$ as $p_b$ approaches $0.5$ but it does not increase dramatically for $p_b \leq 0.3$.

In practice, imperfect detection efficiency and fiber transmission losses can severely reduce $p_b$, thus compromising the success rate of the protocol. In order to circumvent this problem, it is possible to use multiple copies of the resource state $|\psi_{in} \rangle$ and proceed as follows: for each communication interval, if Alice (Bob) detects at least one photon, she (he) assumes that the parity bit is $0$ ($1$).

For our implementation, we use an SPDC-based heralded single-photon source. In each communication interval the source emits $n_e$ photons at different times, of which $n$ are detected. We assume the photon-detection statistics are poissonian. The probability that Alice and Bob together detect $n$ photons is therefore:
\begin{align}\label{eq:poisson}
p(n) = e^{- m} \frac{m^n}{n!},
\end{align}
with $m$ average number of detections.

There are three possible cases in which an error occurs in a given interval: 1) no photon is detected at all, 2) both Alice and Bob detect photons, 3) all photons go to the wrong output. In case 1 Alice (Bob) always infers a value of ``1''(``0'') for the parity bit. The two values are swapped in case 2. Since the parity bit is random, this produces an error in the message bit transmission $50 \%$ of the times. In case 3 the wrong message bit is transferred $100 \%$ of the times. This results in:
\begin{align}\label{eq:biterrorm}
p_b = \frac{p(0)}{2} + \frac{p_{dAB}}{2} + p_{aw},
\end{align}
where $p(0)$, $p_{dAB}$ and $p_{aw}$ are the probabilities of cases 1, 2 and 3, respectively. We have:
\begin{gather}
p_{dAB} = \sum\limits_{n = 2}^{\infty}p(n)\sum\limits_{k=1}^{\infty} {n\choose k} (1 - p_s)^k p_s^{n-k} =\\ \label{eq:pdab}
= 1 + p(0) - \sum\limits_{n = 0}^{\infty} p(n) p_s^n - \sum\limits_{n = 0}^{\infty} p(n) (1 - p_s)^n \nonumber\\
p_{aw} = \sum\limits_{n = 1}^{\infty} p(n) (1 - p_s)^n =\\         \label{eq:paw}
= \sum\limits_{n = 0}^{\infty} p(n) (1 - p_s)^n - p(0), \nonumber
\end{gather}
where $p_s$ is the success probability per photon, that is the probability a photon is detected at the right output.\\
By using the last two equations in expression \ref{eq:biterrorm} and expliciting $p(n)$, we obtain, after some simple passages:
\begin{align}\label{eq:pbmfinal}
p_b = \frac{1}{2}(1 + e^{- m p_s} - e^{- m (1 - p_s)}).
\end{align}
This expression tends to $\frac{1}{2}$ for $m \rightarrow 0$, when the term $\frac{p(0)}{2}$ becomes dominant, and for $m \rightarrow \infty$, when the main contribution to $p_b$ is given by $p_{dAB}$. In between these two regimes, $p_b$ has a minimum at:
\begin{align}\label{eq:minm}
m_{optm} = \frac{1}{2p_s - 1}log(\frac{p_s}{1 - p_s}).
\end{align} 

In some situations, one might wish to optimize the probability both Alice and Bob bits are correctly transferred. An error on the bit-pair transmission occurs any time no photon is detected at all or at least one photon in the encoding interval comes out at the ``wrong'' port. If we call $p_w$ the probability of the latter case, the probability of error on the bit-pair transfer is:
\begin{align}\label{eq:biterrorp}
p_{bpair} = p(0) + p_w 
\end{align}
By expliciting $p(0)$ and $p_w$, we obtain:
\begin{align}\label{eq:biterrorp2}
p_{bpair} = p(0) + \sum_{n=1}^{\infty} p(n) (1 - p^n_s) = \\
 = 1 + e^{-m} - e^{-m(1 - p_s)}. \nonumber
\end{align}
This expression tends to $1$ for $m \rightarrow 0$ and $m \rightarrow \infty$ and has a minimum for:
\begin{align}\label{eq:minp}
m_{optp} = - \frac{log(1 - p_s)}{p_s}.
\end{align}
We note that the distance between $m_{optm}$ and $m_{optp}$ tends to 0 as $p_s$ tends to $1$.

Since the value of $p_s$ for our set-up is approximately $0.95$, we calculate that $m_{optm}$ and $m_{optp}$ are both  around $3$. We reduce the pump power of our source until we obtain globally about $3$ detections per communication interval.

We measure a success probability of the protocol, $1 - p_{b}$, of $0.88 \pm 0.01$ over $10$ sets of $100$ random  pairs of bits. The error is calculated by considering the standard deviation of the probability over the sets and by dividing it by the square root of the number of  sets, thus obtaining an error on the average value. The average number of detections, $m$ and the probability of success per photon, $p_s$, over the sets are respectively $3.34 \pm 0.06$ and $0.935 \pm 0.008$. By inserting these values in  our poissonian model, we calculate a theoretical success probability of $0.88 \pm 0.01$, perfectly compatible with the measured value. We note that, considering $p_s = 0.935$, the maximal theoretical success probability is $0.881$, for $m = 3.061$.

In the same experimental conditions the probability $1 - p_{bpair}$ is measured to be $0.75 \pm 0.02$, with an expected value of $0.77 \pm 0.02$.

We implement the majority-voting error-correction scheme with three and five repetitions per bit pair. In both cases we average over four sets of $100$ bit pairs, for a total of $1200$ and $2000$ iterations, respectively. In the case of three repetitions, we obtain a success probability of $0.93 \pm 0.01$. The expected value in this case is $0.969 \pm 0.004$, considering that we measure $m = 2.62 \pm 0.05$ and $p_s = 0.948 \pm 0.006$. For the case of five repetitions, the measured success probability results to be $1.00 \pm 0.01$ while the expected value is $0.995 \pm 0.001$, with the measurement results $m = 3.736 \pm 0.004$ and $p_s = 0.960 \pm 0.004$. All the measured values are compatible with the expected ones, thus confirming the validity of our theoretical model.

\section*{Appendix E: Security analysis of the anonymous communication protocol}
The anonymous communication protocol is information-theoretically secure under noise, losses, and repetitive transmission.
Without loss of generality this analysis is done for communication from Alice to Bob only.
The other direction of communication is covered by symmetry.
The analysis, overall performed for {\em noisy\/} protocol executions and potential multi-photon emissions, covers three steps: 
First, we show security in the case with {\em losses\/}, then we model the {\em probabilistic photon source\/} as has been used in the experimental setup and show security in that case, and finally we show security for the {\em repetition code\/} where multiple rounds are used in order to reliably transmit a single bit.
Once we achieve secure and reliable transmission of a single bit, multi-bit messages can be transmitted securely and reliably by repeating the protocol. 
Integrity and authenticity can be achieved by the use of message authentication, which, however, needs some pre-shared randomness between the parties.

We prove security in the ideal-real model (also known as the simulator model).
The {\em ideal\/} cryptographic primitive we want to implement is a {\em secure} transmission of a bit~$x$ from Alice to Bob.
An adversary, in this ideal scenario, does not learn anything about~$x$; she or he only learns that the primitive has been used.
The {\em real\/} scenario, then again, is the protocol as explained above.
If the adversary cannot {\em distinguish\/} a real protocol execution from the ideal one, then the real protocol is called secure.
In other words, if the adversary can {\em simulate\/} the data he or she sees in a real execution from the ideal primitive (no access to the inputs), then the real protocol is secure.
This notion is covered by the following definition.
\begin{definition}[Approximate security]
	We call the protocol~{\em $(1-\delta)$-secure\/} if there exists some density operator~$\tau$ such that for every input~$x$ from Alice we have \\
	$\Delta(\tau,\sigma)\leq \delta$,
	where~$\Delta(\cdot,\cdot)$ is the trace distance and~$\sigma$ is the density operator a potential eavesdropper has access to.
\end{definition}
Note that a~$(1-\delta)$-secure protocol implies that the eavesdropper can guess Alice's bit~$x$ not better than with probability~$1/2+\delta$.
This follows, because~$\delta\geq\Delta(\tau,\sigma)\geq\sum_{x'}|P(x')-1/2|/2=(2P(x)-1)/2$, where, given~$\tau$, the probability of guessing~$x$ is~$1/2$, and given~$\sigma$, the probability of guessing~$x$ is~$P(x)\geq 1/2$.

Throughout the analysis we will consider {\em noisy\/} and {\em lossy\/} resource states, which motivates the statement of the following definition.
\begin{definition}[Security parameter (noise) and loss]
	A resource state~$\rho^\nu$ has {\em security parameter~$\nu$\/} if
	\begin{align}
		\min_{
			\substack{
			n,\ket{\psi}\in\operatorname{span}(
			\{
				(\hat a^\dagger)^k(\hat b^\dagger)^{n-k}\ket{0} \,|\, 0\leq k \leq n
			\}):\\
			\langle{\psi}\,|\,\psi\rangle=1
			}
		}
		\Delta\left( 
		\ket\psi\bra\psi,\rho^\nu
		\right)
		\leq\nu
		\,.
	\end{align}

	A resource state~$\rho^{\nu,\lambda}$ {\em with~$\lambda$ loss\/} is~$(1-\lambda)\rho^{\nu}+\lambda\ket{0}\bra{0}$, where~$\ket{0}\bra{0}$ is the vacuum state.
\end{definition}
The security parameter reflects how much an adversary might learn about the bit~$x$.
Note that this does not depend on the number of photons generated; hence, the protocol is resistant against multi-photon emissions.
As will be shown in the proof, if the resource state lies in the span of having~$k$ photons with Alice and~$n-k$ with Bob (for some fixed~$n$), then the protocol is perfectly secure.
The definition for lossy resource states is done in agreement with the theoretical letter \cite{Bori}.
For the analysis, it is helpful to consider the pure version of the protocol: All but the last operations are unitary.
In the last step, Bob measures his registers, which consist of a space containing any number of particles from $0$ to $n$ and his random setting, with some appropriate observable.
To generate the random one-time pad, Bob uses the~$\ket +=(\ket 0+\ket 1)/\sqrt 2$ state and encodes that setting with a unitary operation.
The randomness is generated by measuring that register.

\begin{theorem}
	For given~$\nu$ and~$\lambda$, the protocol is~$(1-(1-\lambda)\nu)$-secure.
\end{theorem}
\begin{proof}
	Let~$\rho^{\nu,\lambda}$ be the resource state, and~$\sigma^{\nu,\lambda}$ be the state an adversary has access to.
	We first use the triangle inequality of the trace distance:
	\begin{align}
		\Delta\left( 
		\tau,
		\sigma^{\nu,\lambda}
		\right)
		\leq
		\Delta\left( 
		\tau,
		\sigma^{0,\lambda}
		\right)
		+
		\Delta\left( 
		\sigma^{0,\lambda}
		,
		\sigma^{\nu,\lambda}
		\right)
		\,.
	\end{align}
	There exists some~$\tau$ that makes the first expression on the right equal to~$0$: The reduced state~$\sigma^{0,\lambda}$ after tracing over Alice's input and Bob's randomness is independent of~$x$.
	This is shown in the following calculation, where~$\ket\psi$ is the~$n$-photon state that minimizes the security parameter~$\nu$.
	Having~$\ket\psi$ as resource state, the initial state of the protocol is
	\begin{align}
		\ket\psi\otimes\ket +\otimes\ket x
		=
		\left( 
			\sum_{k=0}^n
			\alpha_k
			\left(\hat a^\dagger\right)^k
			\left(\hat b^\dagger\right)^{n-k}
			\ket 0
		\right)
		\otimes\ket +\otimes \ket x
		\,,
	\end{align}
where the last register is Alice's message register and the one before Bob's one-time-pad register.
	After the parties encode their bits, the state is transformed to
	\begin{gather}
	\frac{1}{\sqrt{2}}
		 \left(
			\sum_{k=0}^n
			\alpha_k
			(-1)^{kx}\left(\hat a^{\dagger} \right)^k
			\left(\hat b^{\dagger}\right)^{n-k}
			\ket 0
			\right)
			\otimes\ket 0 \otimes \ket x +\\
+	\frac{1}{\sqrt{2}}
		    \left(
			\sum_{k=0}^n
			\alpha_k
			(-1)^{kx+(n-k)}
			\left(\hat a^{\dagger} \right)^k
			\left(\hat b^{\dagger}\right)^{n-k}
			\ket 0
			\right) \otimes \ket 1
		\otimes \ket x. \nonumber
	\end{gather}

By tracing over the message space and the randomness register, we obtain~$\sigma^{0,\lambda}$, which is
	\begin{gather}
		\frac{1}{2}\sum_{k,k'=0}^n
		\left( 
		(-1)^{(k+k')x}
		+
		(-1)^{(k+k')x+2n-k-k'}
		\right)
		\alpha_k \alpha_{k'}^*...\\ \nonumber
		...\left(\hat a^\dagger \right)^k
		\left(\hat b^\dagger \right)^{n-k}
		\ket 0
		\bra 0
		\left(\hat b \right)^{n-k'}
		\left(\hat a \right)^{k'} =\\ \nonumber
		=
		\sum_{\substack{k,k'=0\\k+k'\text{ is even}}}
		\alpha_k\alpha_{k'}^*
		\left(\hat a^\dagger \right)^k
		\left(\hat b^\dagger \right)^{n-k}
		\ket 0
		\bra 0
		\left(\hat b \right)^{n-k'}
		\left(\hat a \right)^{k'}
		\,.
	\end{gather}
The second expression on the right, then again, can be bounded by using the fact that the trace distances does not increase under any completely-positive trace-preserving map and by using the strong-convexity property:
	\begin{gather}
		\Delta\left( 
		\sigma^{0,\lambda}
		,
		\sigma^{\nu,\lambda}
		\right)
		\leq
		\Delta\left( 
		\rho^{0,\lambda}
		,
		\rho^{\nu,\lambda}
		\right) \leq\\
		\leq
		(1-\lambda)\Delta\left( \ket\psi\bra\psi,\rho^{\nu,0} \right)
		+ \lambda\Delta\left( \rho^{0,1},\rho^{\nu,1} \right)
		\leq
		(1-\lambda)\nu
		\,. \nonumber
	\end{gather}
The term~$\Delta( \rho^{0,1},\rho^{\nu,1} )$ is~$0$: Both density operators represent the vacuum state.
\end{proof}

Note that if the resource state is a mixture of resource states with different photon numbers, then the protocol remains secure, unless the purification registers are leaked to the eavesdropper.

Now, let us discuss the security for the protocol as it is implemented.
The experimental setup uses a probabilistic photon source that generates~$k$ particles, where that number is Poisson distributed with mean value~$m$.
\begin{theorem}
	If the single-particle protocol is~$(1-\delta)$-secure
	then the use of a probabilistic photon source with mean value~$m$ renders the protocol~$(1-m\delta)$-secure.
\end{theorem}
\begin{proof}
	Let~$\rho'$ be the single-particle resource state that makes the single-particle protocol~$(1-\delta)$-secure, and let~$\tau$ be the {\em simulated\/} view of the eavesdropper.
	The resource state generated by the probabilistic photon source is
	\begin{align}
		\rho'_\text{PPS}=
		\sum_{k=0}^\infty p(k) \rho'^{\otimes k}
		\,.
	\end{align}
	A potential eavesdropper, in this case, has access to the reduced state~$\sigma'_\text{PPS}=\sum_{k=0}^\infty p(k)\bigotimes_{i=0}^k\sigma'_{i,k}$, where~$\sigma'_{i,k}$ is the state of the~$i$-th particle in the case~$k$ particles have been generated.
	Define by~$\tau_\text{PPS}$ the state~$\sum_{k=0}^\infty p(k)\tau^{\otimes k}$.
	Due to strong-convexity and subadditivity we have
	\begin{align}
		\Delta\left( 
		\tau_\text{PPS},
		\sigma'_\text{PPS}
		\right)
		\leq
		\sum_{k=0}^\infty
		p(k)
		\Delta\left( 
		\tau^{\otimes k},
		\bigotimes_{i=0}^k
		\sigma'_{i,k}
		\right)
		\leq
		\sum_{k=0}^\infty
		p(k)
		k\delta
		=m\delta
		\,.
	\end{align}
\end{proof}

Finally, we discuss the repetition code:~$N$ (odd) successive rounds are performed where Alice always transmits the same bit~$x$ and Bob uses the majority of the detected values as his guess --- a form of error correction.
\begin{theorem}
	If the protocol is~$(1-\delta)$-secure
	then the repetition code with~$N$ repetitions makes the protocol~$(1-N\delta)$-secure.
\end{theorem}
\begin{proof}
	The proof is analogous to the last.
	Due to the security of the single-run protocol there exists some~$\tau$, independent of~$x$, such that $\Delta\left(\tau,\sigma\right) \leq \delta$, where~$\sigma$ is the state the eavesdropper has access to.
	Denote by~$\sigma_i$ the~$i$-th state the eavesdropper has access to.
	Then
	\begin{align}
		\Delta\left( 
		\tau^{\otimes N},
		\bigotimes_{i=1}^N\sigma_i
		\right)
		\leq N\delta
		\,.
	\end{align}
\end{proof}

\end{document}